\documentclass[singlespacing]{elsart}

 \usepackage{epsfig}

\usepackage{amssymb}

\def\prl{{\sl Phys. Rev. Lett.}\ }
\def\prb{{\sl Phys. Rev. B}\ }

\begin{document}
\begin{frontmatter}


\title{Superconducting islands, phase fluctuations and the superconductor-insulator transition }

 \author[bgu]{Yonatan Dubi}, 
\author[bgu,ilse]{Yigal Meir} and 
 \author[bgu,ilse,france]{Yshai Avishai}
 \address[bgu]{Physics Department, Ben-Gurion University, Beer Sheva 84105, Israel}
 \address[ilse]{The Ilse Katz Center for Meso- and Nano-scale Science and
Technology, Ben-Gurion University, Beer Sheva 84105, Israel }
\address[france]{SPHT, CEA, F91191 Gif sur Yvette, France}


\date{\today}

\begin{abstract}
Properties of disordered thin films are discussed based on the viewpoint that
superconducting islands are formed in the system. These lead to 
 superconducting correlations confined in space, which are 
known to form spontaneously in thin films. 
Application of a perpendicular magnetic field 
can drive the system from the superconducting state (characterized by phase-rigidity between 
the sample edges) to an insulating state in which there are no phase-correlations between the 
edges of the system. On the insulating side the existence of superconducting 
islands leads to a non-monotonic magnetoresistance. Several other features seen in 
experiment are explained. 
\end{abstract}


\end{frontmatter}


\section{Introduction}
Superconductivity (SC) in disordered thin films has been a subject of intense study for more than a 
decade \cite{Goldman_review}. Nevertheless, even 
elucidation of one of the most fundamental property of these systems, namely the 
superconductor-insulator transition (SIT), remains a puzzle. Specifically, it is still unclear 
whether this is a truely quantum phase-transition, what is the role of the magnetic 
field in the transition, and even its universality class remains undetermined. Another 
profound feature which is still in debate is the non-monotonic magneto-resistance (MR), which 
in some systems can reach several orders of magnitude \cite{Murty}. The non-monotonic MR is 
accompanied by a unique temperature, magnetic field and disorder dependence of the resistance 
\cite{Murty,Kapitulnik} .
On the theoretical front, the adequacy of treating the system in terms of 
strictly bosonic excitations (so called "dirty boson" models \cite{dirtybosons}) 
is still questionable.

We adopt a perspective within which the system is composed of SC islands, a structure implying local SC correlations. 
 The notion of SC islands appeared nearly a decade ago in two contexts. Galitski 
and Larkin \cite{Galitzki} suggested that in a strongly disordered system the SC order parameter
 ( and hence
also $T_c$ and $H_c$)  fluctuate in space. As a result, at a given temperature and magnetic field (close to  $H_c$ of the clean system) 
there will be areas in the sample where the local critical field exceeds the external field. Therefore these domains  still display 
SC correlations. Ghosal et.al. \cite{Ghosal} used a locally self-consistent solution of the Bogoliubov-de-Gennes equations to 
show that even in the presence of extremely strong disorder there remain regions in space where the SC order parameter is 
finite, surrounded by regions of vanishingly small order parameter (and hence dubbed SC islands). 

As it is shown below, the concept of SC islands might explain  numerous
experimental observations.  In section 2 we describe the nature of the SIT as it emerges from the existence of 
SC islands with fluctuating phases. In section 3 we show how SC islands lead to non-monotonic MR and explain its dependence on 
temperature and magnetic field. Section 4 is devoted to a summary and outlook.

\section{From Superconducting islands to the superconductor -- insulator transition}
In this section we describe the nature of the SIT, based on our previous investigations
\cite{us_Nature}. 
On the SC side of the transition, the system has a complex SC order parameter, and its amplitude strongly fluctuates in 
space. When a perpendicular magnetic field is turned on and then monotonically increased, it begins to penetrate the sample in the form of 
disordered vortices \cite{us_PRB}. 
As a consequence, there are regions in which the SC order parameter vanishes, and SC islands 
are formed. However, these islands are still interconnected via a Josephson coupling (JC), and hence their phases are well 
correlated. This means that Cooper-pairs can coherently traverse through the sample and carry super-current, i.e. the system 
is in a macroscopic SC state. 

When the JC between two islands becomes smaller than the temperature, 
thermal phase fluctuations overcome the phase-locking 
induced by the JC, and the islands become separate (in terms of SC wave functions). Now, as the magnetic field is increased the JC 
between the different islands decreases. However, as long as there is a phase-stiff path between the edges of the sample (that 
is a path of islands with JC larger than temperature) the system will maintain its SC nature. At a certain magnetic field $B_c$, the 
phase-stiff path is broken and the system is no longer in a SC state. Since it is a highly disordered two-dimensional system, 
one expects that it it will become insulating, as indeed observed.

An immediate consequence of the above scenario is that SC correlations still survive on the insulating 
side of the transition. This idea is corroborated by a number of experiments. For instance, 
measurements of the AC conductance \cite{Crane} show that the superfluid stiffness 
remains finite well above the transition. Another example is magnetic-field induced 
conductance oscillations with $2 \phi_0$ period observed in a sample with a fabricated 
nano-hole lattice \cite{Valles}, indicating the presence of Cooper pairs. 

\begin{figure}[h!] 
\centering
\includegraphics[width=8 truecm]{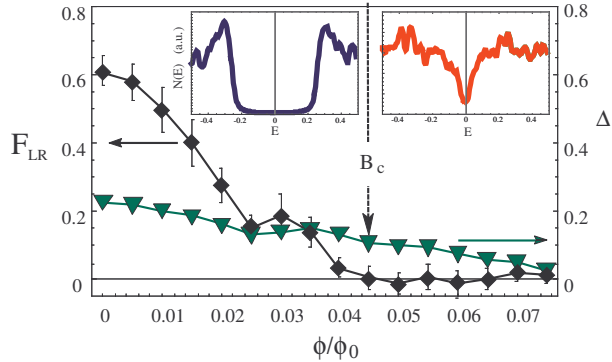} 
\centering
\caption{\footnotesize Phase correlations between the two sides of the system $F_{LR}$ (diamonds) 
and the SC order parameter amplitude $\bar{|\Delta|}$ (triangles) as a function of magnetic field. At a 
certain magnetic field $B_c$ phase-correlations vanish, indicating the SIT. However, 
$\bar{|\Delta|}$ is still finite after this point, indicating the existence of SC islands. Insets : the density of states for vanishing magnetic field (left inset) and just above the transition (right inset), demonstrating the appearance of a "pseudogap" above the transition.   }
\label{PhaseCorr} \end{figure}

To support the above picture we have performed numerical calculations of phase 
correlations between the two edges of a highly disordered superconducting thin film. We used  
a Monte-Carlo scheme \cite{Mayr} on-top of a self-consistent solution of the Bogoliubov-de-Gennes 
mean-field equations to obtain both the order parameter amplitude and phase, and calculated the 
correlation function $F_{LR}=\langle \cos (\delta \theta_i - \delta \theta_j) \rangle$ (where 
$\delta \theta_i$ is the shift of the order parameter phase from its mean-field value)
as a 
function of magnetic field (in terms of the flux per plaquette, $\phi/\phi_0$). Here 
$i(j)$ are lattice sites on the left (right) edge of the sample. Details of the 
calculations are described elsewhere \cite{us_Nature}. In Fig.~\ref{PhaseCorr} we show $F_{LR}$ (diamonds) 
as a function of magnetic field (or flux per plaquette) for a single disorder realization of a 
system of size $20 \times 5$, with 
strong disorder $W/t=1.4$ (where $t$ is the hopping integral), attractive interaction strength 
$U/t=2$, average electron density (per site) $\langle n \rangle =0.92$ and temperature $T=0.04$. In addition we plot the average 
value of the SC order parameter amplitude $\bar{|\Delta|}$ (triangles), for the same magnetic field. As can be
seen, at a certain (disorder and density dependent) magnetic field the phase correlation vanish, and 
hence this value corresponds to $B_c$. However, the order parameter amplitude is finite beyond this 
point and vanishes only at higher fields. This can also be seen from the density of states (DOS), 
plotted in the inset of Fig.~\ref{PhaseCorr} for two values of the magnetic field. The left inset 
displays the results for vanishing magnetic field, 
where a clear BCS-like DOS is observed. The right inset pertains to magnetic field of 
$\phi/\phi_0=0.06$, that is above the SIT. However, since there are still SC correlations a 
pseudogap develops. This may indicate that our picture is  relevant also to the physics of high Tc 
superconductors \cite{Alvarez}. We point that similar results were obtained from tunnelig measurments \cite{valles}, 
where the broadened BCS peak was accounted for by SC order parameter amplitude fluctuations. 

Although  the transition is described in terms of thermal fluctuations, this 
description does not rule out the possibility that the transition is quantum in nature. All one needs to 
do is to replace thermal fluctuations with quantum fluctuations and temperature with an energy 
scale associated with quantum fluctuations. In fact, recent measurements \cite{Crane,Aubin} 
indicate that the transition crosses smoothly from a thermal to a quantum phase transition, the phenomenology of the two classes being very similar. 

\section{Superconducting islands and non-monotonic magneto-resistance}
In a recent set of experiments, Sambandamurthy et.al. \cite{Murty} showed that the magneto-resistance
is non-monotonic above the SIT and develops a peak at a certain magnetic field $B_\mathrm{max}$. Surprisingly, the resistance at the peak can 
be several orders of magnitude larger than the resistance at the transition (which is found {\sl 
not} to be universal, see e.g. \cite{Hebard,kapitulnik1}). Perhaps even more surprising is the fact that with increasing the magnetic 
field beyond $B_\mathrm{max}$ the resistance drops several orders of magnitude and comes close to 
its value at the SIT. While non-monotonicity in the MR was observed more than a decade ago \cite{Non-Mon}, there it was a miniscule effect. Here, however, this effect 
is huge and cannot be overlooked. In addition, the resistance develops an activation dependence on 
temperature, which vanishes at low enough temperatures \cite{Murty}.
An experimental  study of MR for different values of 
disorder \cite{Kapitulnik} shows that with increasing disorder the value of the resistance 
at the MR peak increases, and at the same time the values of both $B_c$ and $B_\mathrm{max}$ 
decreases. 

How all these phenomena can be explained in terms of the formation of SC 
islands \cite{us_MR} is discussed below. The model is based on three
assumptions. The first is that disorder induces the formation of SC
islands due to fluctuations in the amplitude of the SC
order parameter. The second
assumption is that as the magnetic field is increased, the
concentration and size of these SC islands  decrease. 
This does not mean, however, that the physical 
size of the islands (that is, the spatial extent to which the order parameter amplitude is finite) decreases (although it might), but rather 
that islands become disconnected in the sense of 
phase-fluctuations described in the previous section. 
The third assumption is that the SC islands 
have a charging energy, and thus, a Cooper pair entering a SC island (via an
Andreev tunneling process) has to overcome it.
This charging energy is expected to be inversely proportional to
the island size, and thus to increase with increasing magnetic
field. All three assumptions have been corroborated 
by numerical calculations described elsewhere \cite{us_PRB}.

Consider now such a system in the strong magnetic field
regime, $B>>B_\mathrm{max}$. Due to the strong magnetic field
the SC islands are small and have a large charging energy. Now the electrons can traverse 
through the system in two types of trajectories: those
which follow normal areas of the sample and do not cross the SC islands ("normal paths") 
and those in which an electron tunnels into a
SC island via the Andreev channel ("island paths"). 
The resistance of the normal paths, $R_N$, has some value
(which may depend on e.g. length, temperature, disorder etc.) and is
assumed to be weakly affected by magnetic field. Due to 
Coulomb blockade, transport through the "island paths" is thermally
activated, and hence the resistance of the island paths is of the
form $R_I \sim \exp (E_{c}/T)$, where $E_c$ is the typical charging energy
of the island. If $E_c$ is large then the main contribution to the
conductance is due to transport along the normal paths. Consistent
with experiment, the MR in this regime is small.
\par As the magnetic field is decreased (but still in
the regime $B>B_\mathrm{max}$), more SC islands are created and
their size increases, but they are still small enough such that
transport along normal paths is favourable. However, some paths which were
normal at higher fields now
become island paths and hence unavailable for electron transport. Thus, the effective
 density of normal electrons which contribute to the resistance diminishes, resulting
in a negative MR. Eventually, at a certain magnetic field $B=B_\mathrm{max}$
 some SC islands are large  enough so that their charging
energy is small and the resistance through them is comparable to
the resistance through normal paths, i.e. 
$
R_N \approx R_I 
$. At this point the resistance
reaches its maximum value, since as the magnetic field is further decreased
the SC islands are so large that transport through them is
always preferred over transport through normal paths. An increase in number and size
of the SC islands will thus result in a decrease in the resistance. At the critical 
field $B_{c}$ a phase-stiff path percolates through
the system, resulting in the SIT.

The above model was encoded into a numerical calculation using a lattice network model of normal resistors, 
SC resistors and insulating resistors (to mimic the Coulomb blockade) \cite{us_MR}. Resistance of a normal resistor 
connecting two sites on the lattice is given by $
R_{ij} =R_0 \exp \left( \frac{2 r_{ij}}{ \xi_{loc}}+\frac{|\e_{i}|+|\e_{j}|+|\e_{i}-\e_{j}|}{2kT} \right),
 $ where $R_0$ is a constant, $r_{ij}$ is the distance between sites $i$ and $j$,
$\xi_{loc}$ is the localization length, $\e_{i}$ is the energy of
the $i$-th site measured from the chemical potential (taken from a
uniform distribution $[-W/2,W/2] $) and $T$ is the temperature. The resistance between two (neighboring) SC sites is taken to be very small compared with that of a normal resistor,
but still not zero and temperature dependent, in such a way that it vanishes as $T\rightarrow 0$ (distant SC sites
are disconnected). The resistance between a
normal site and a SC site (i.e. an insulating resistor) is taken to be $ R_{NS} \propto \exp
(E_{c}/kT), $ where $E_c$ is the charging energy of the
island.

\begin{figure}[h!] 
\centering
\includegraphics[width=8truecm]{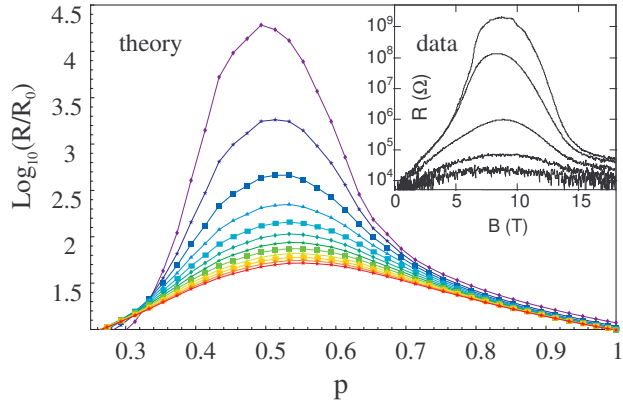} 
\centering
\caption{\footnotesize Magneto-resistance as a function of concentration of SC islands for different temperatures (taken from Ref.~\cite{us_MR}), showing good quantitative resemblance between the theory and experimental data (inset, taken from Ref.~\cite{Murty}) is evident.   }
\label{MR} \end{figure}

In Fig.~\ref{MR} we
plot the resistance as a function of the concentration of the SC islands (which corresponds to the 
magnetic field) for different values of temperature. The quantitative resemblance to the experimental data (inset of Fig.~\ref{MR}) is 
self-evident. In addition, this simple phenomenological theory is capable of accounting for the 
breakdown of activation behaviour at low temperatures, the dependence of disorder and other 
experimental observations.    

\section{Summary and outlook}
In this contribution we have demonstrated that many of the properties of disorder SC thin films may 
be accounted for by the formation of SC islands. Specifically, we have demonstrated that 
phase-fluctuations between different islands lead to the SIT, and that above the transition formation of SC 
islands leads to a non-monotonic magneto-resistance. 

Still, there are many puzzles left in elucidating the physics of these systems. For instance, an unusual disorder and magnetic-field dependent anisotropy in the magneto-resistance was 
recently measured \cite{angular} and is up-to-date unexplained, although it seems that a percolation theory similar to that presented here may 
account for it \cite{Yigal}. Even more challenging is the appearance of a seemingly correlated \cite{Shahar2} and perhaps metallic 
state at high magnetic field \cite{Baturina}. A final example is the fact that some materials seem 
to exhibit a metallic intermediate state at the SIT while others do not \cite{Baturina2}. 
These and other questions are still waiting to be resolved.

\end{document}